\documentclass[preprint,12pt]{elsarticle}
\usepackage{subcaption}
\usepackage{amssymb}
\usepackage{amsmath}
\usepackage{amsthm}
\journal{Journal of Energy Storage}

\begin{document}

\begin{frontmatter}

\title{State Estimation for Parallel-Connected Batteries via Inverse Dynamic Modeling}

\author{Hannah Lee$^{a}$, Casey Casten$^{a}$, Hosam K. Fathy$^{b}$} 

\affiliation{Research intern, University of Maryland.}
\affiliation{Professor of Mechanical Engineering at University of Maryland.}

\begin{abstract}
This paper examines the problem of estimating the states, including state of charge, of battery cells connected in parallel. Previous research highlights the importance of this problem, and presents multiple approaches for solving it. Algorithm scalability and observability analysis can both be challenging, particularly because the underlying pack dynamics are governed by differential algebraic equations. Our work addresses these challenges from a novel perspective that begins by inverting the causality of parallel pack dynamics, which breaks the pack model's underlying algebraic loop. This simplifies observability analysis and observer design significantly, leading to three novel contributions. First, the paper derives mathematical conditions for state observability that apply regardless of the number of battery cells and the order of their individual dynamics. Second, the paper presents an approach for grouping battery cells such that their lumped dynamics are observable. Finally, the paper presents a novel pack state estimator that achieves computational tractability by employing inverse dynamic modeling. We conclude by presenting a Monte Carlo simulation study of this estimator using experimentally-parameterized models of two battery chemistries. The simulation results highlight the computational benefits of both the clustering strategy and inverse dynamics approach for state estimation.
\end{abstract}

\begin{highlights}
\item Inverse dynamic modeling simplifies observability analysis for parallel-connected battery cells. 
\item Potentiostatic current relaxation eigenvalues can be used to group battery cells for state of charge estimation. 
\item Inverse dynamic modeling leads to computationally efficient state of charge estimation. 
\end{highlights}


\end{frontmatter}


\pagebreak
\section{Introduction}
\label{sec1}

\begin{figure}[h]
    \centering
    \includegraphics[width=0.8\textwidth]{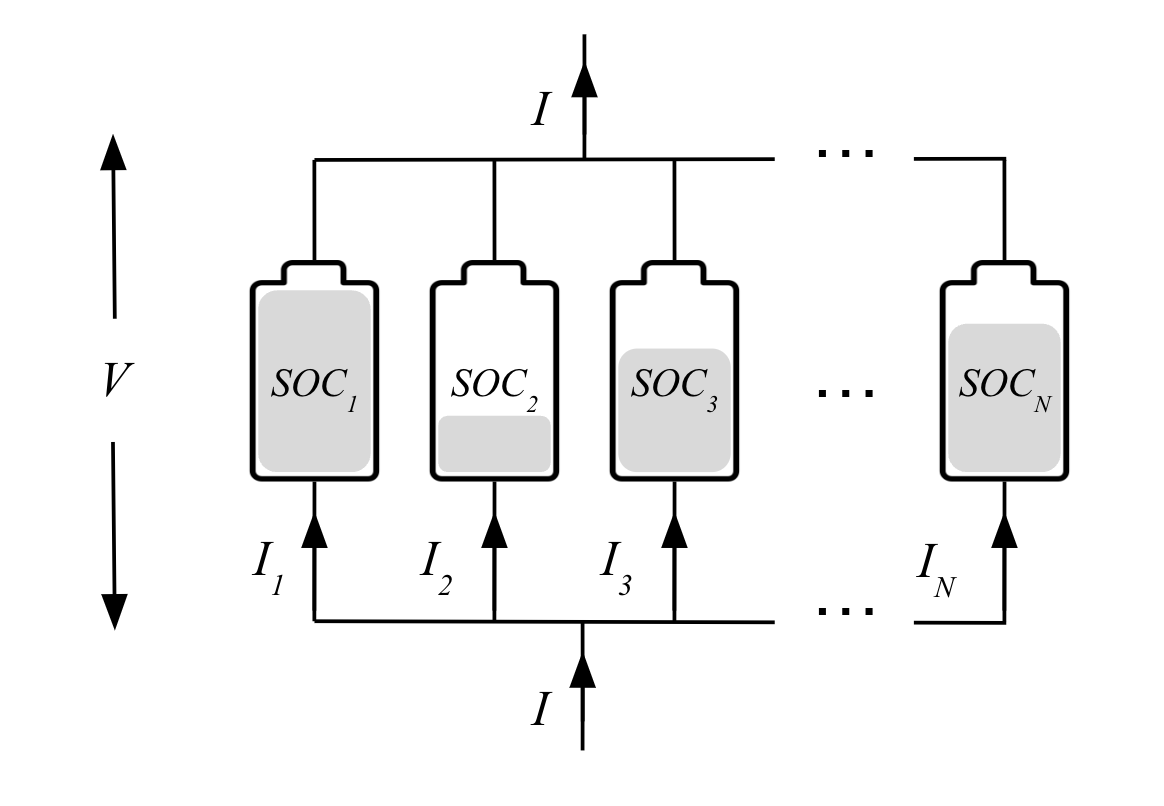}
    \caption{Configuration of a parallel-connected battery pack.}
    \label{fig:parallelPack}
\end{figure}

This paper examines the problem of estimating the state variables, including state of charge (SOC), for battery cells connected in parallel, as shown in Fig.~\ref{fig:parallelPack}. We demonstrate our findings using two lithium-ion battery examples, but the underlying mathematical framework is applicable to other chemistries. Motivation for this work is twofold: 

\begin{itemize}
    \item First, there is a growing need for large battery packs in applications such as electric vehicles and grid energy storage~\cite{ding2019automotive}. Thousands of cells are typically connected in series and/or parallel in these packs. Series connections increase pack voltage, while parallel connections improve overall pack reliability through redundancy.
    \item Second, inevitable discrepancies in parameters such as charge capacity and internal resistance may compromise safety and efficiency in parallel-connected battery packs. Specifically, these discrepancies trigger imbalanced current distributions among the different cells. This can accelerate cell aging compared to uniform packs~\cite{shi2016imbalance} and dissipate energy as heat through cell-to-cell leakage currents during rest. Careful battery selection can reduce cell-to-cell heterogeneity at the beginning of life~\cite{kim2013screening}, but heterogeneity can grow with battery pack use and age. 
\end{itemize}

The above facts create a need for battery management systems (BMSs) that can track the imbalance between the states, including SOC, of parallel cells. Such tracking can help minimize cell degradation and energy loss by enabling heterogeneity-aware management and diagnostics. 

There is a rich literature on pack-level state estimation. Much of this literature focuses on series connections, but the underlying ideas can often be applied to parallel connections as well. Pack-level estimators often build on existing single-cell state estimation algorithms, such as Kalman filtering~\cite{plett2004EKF,plett2006SPKF} or data-driven methods~\cite{chen2016ml,charkhgard2010ml,how2020ml,hong2020ml}. However, there is a growing understanding that representing a battery pack as a single cell without accounting for heterogeneity can lead to poor estimation~\cite{wang2015capacity}. This raises an important question: \textit{how can one account for cell-to-cell heterogeneity in pack-level state estimation?}

There are various approaches to the above question. These include estimation using a lumped model tailored to fit a heterogeneous parallel-connected pack's dynamics~\cite{wang2021pack}, estimation of current distributions using ``representative cells" for a parallel pack's different branches~\cite{yu2023repcell}, and the detection of faulty modules as part of the estimation process~\cite{zhou2021average}. Other approaches account for heterogeneity by estimating the interval of SOCs among parallel cells~\cite{zhang2021interval}, or the extreme SOCs~\cite{zhong2014b1andb2}, or the minimum cell SOC during discharge~\cite{wang2022mini}. Another approach is to partition pack dynamics in the frequency domain into a mean SOC model plus individual cell deviations~\cite{plett2009bardelta, dai2012mean,zheng2013mean,sun2015mean,zheng2018mean}. The literature examines this approach for series connections, but it can potentially be employed for parallel connections as well.

Regardless of the approach, state variables (including SOC) can only be estimated if they are \textit{observable} (i.e., if the estimation problem is solvable). Rausch et al. show that local SOCs are recoverable as long as no two cells' parameters and initial conditions are identical~\cite{rausch2013observe}. Similarly, Zhang et al. find that all cells must have nonzero slopes of open circuit voltage (OCV) with respect to SOC \cite{zhang2021observe} and that no two cells can share the same voltage relaxation constant or OCV function \cite{zhang2020observe}.

The above literature provides important foundations for this paper, but at least three challenges remain. First, the literature relies on battery models with \textit{forward causality}, where current is the input variable and voltage is the output variable. For parallel connections, unlike series connections, this furnishes differential algebraic equation (DAE) models of multi-cell dynamics. This increases the mathematical complexity of pack-level observability analysis. Second, there is a need for a simple criterion for grouping cells in a large pack such that their resulting lumped dynamics are kept observable. Third, there continues to be a need for computationally tractable online state estimators for parallel-connected cells, which is critical given the computational complexity associated with DAE models. There exist approaches to solving these DAEs efficiently, such as waveform relaxation  \cite{saccani2022computationally}. However, this problem could still benefit from additional simplifications that improve estimation tractability.   

This paper presents a simple but novel mathematical framework that addresses all of the above gaps. Namely, we begin by presenting an \textit{inverse causality} model of cell dynamics, where voltage is the input variable and current is the output variable. Practically speaking, this change of perspective has no negative consequences: regardless of whether a BMS directly controls a pack's current or voltage, both quantities are typically assumed to be measured for the purpose of SOC estimation. Therefore, any \textit{online estimator} (as opposed to \textit{controller}) that relies on these measurements is feasible, regardless of its internal causality. A key benefit of causality inversion is that it eliminates the algebraic loop associated with the splitting of current among parallel cells. This results in a system of explicit ordinary differential equations (ODEs), as opposed to DAEs. This simplification leads to the paper's three novel contributions to the literature: 

\begin{itemize}
    \item First, the paper derives mathematical conditions for local SOC observability in parallel-connected cell systems. These conditions encompass all pertinent findings from the earlier literature, albeit with a much simpler derivation. They can also be easily extended to higher-order cell models. This generalizability is important for both lithium-ion batteries and other emerging chemistries where higher-order dynamics may be important \cite{feng2021sodium,rabab2023sodium}.  
    \item Second, the paper presents a simple criterion for clustering cells such that their lumped dynamics are observable. 
    \item Third, the paper designs a Kalman filter for clustered SOC estimation using the inverse causality approach, and demonstrates its accuracy using a Monte Carlo simulation for experimentally parameterized models of two different lithium-ion battery chemistries. 
\end{itemize}

The remainder of this paper is organized as follows: Section 2 establishes a linearized inverse causality model for parallel cells. Section 3 conducts a local SOC observability analysis on this model. Section 4 applies this analysis to the design of a steady-state Kalman filter based on inverse causality modeling and a simple clustering criterion. Section 5 verifies the filter using a Monte Carlo simulation. Section 6 summarizes the paper's conclusions. 


\section{Parallel-Connected Pack Modeling}
\label{sec2}

\subsection{Model Structure}
\label{subsec1}

Fig.~\ref{fig:parallelPack} shows the structure of a group of $N$ parallel-connected cells. Multiple such groups can be connected in series to form a higher-voltage battery pack. The focus in this paper is on SOC estimation at the parallel-connected level, with the understanding that solving this problem is a key step towards SOC estimation in larger series connections of such groups. The total input current, $I(t)$, is typically controlled by the BMS. Moreover, the output voltage, $V(t)$, is typically measured. Motivated by this architecture, the literature typically presents \textit{forward causality} models of parallel-connected battery cells, as shown below. 

\subsection{Forward Causality Model}
\label{subsec2}

\begin{figure}[h]
\centering
\includegraphics[scale = 0.41]{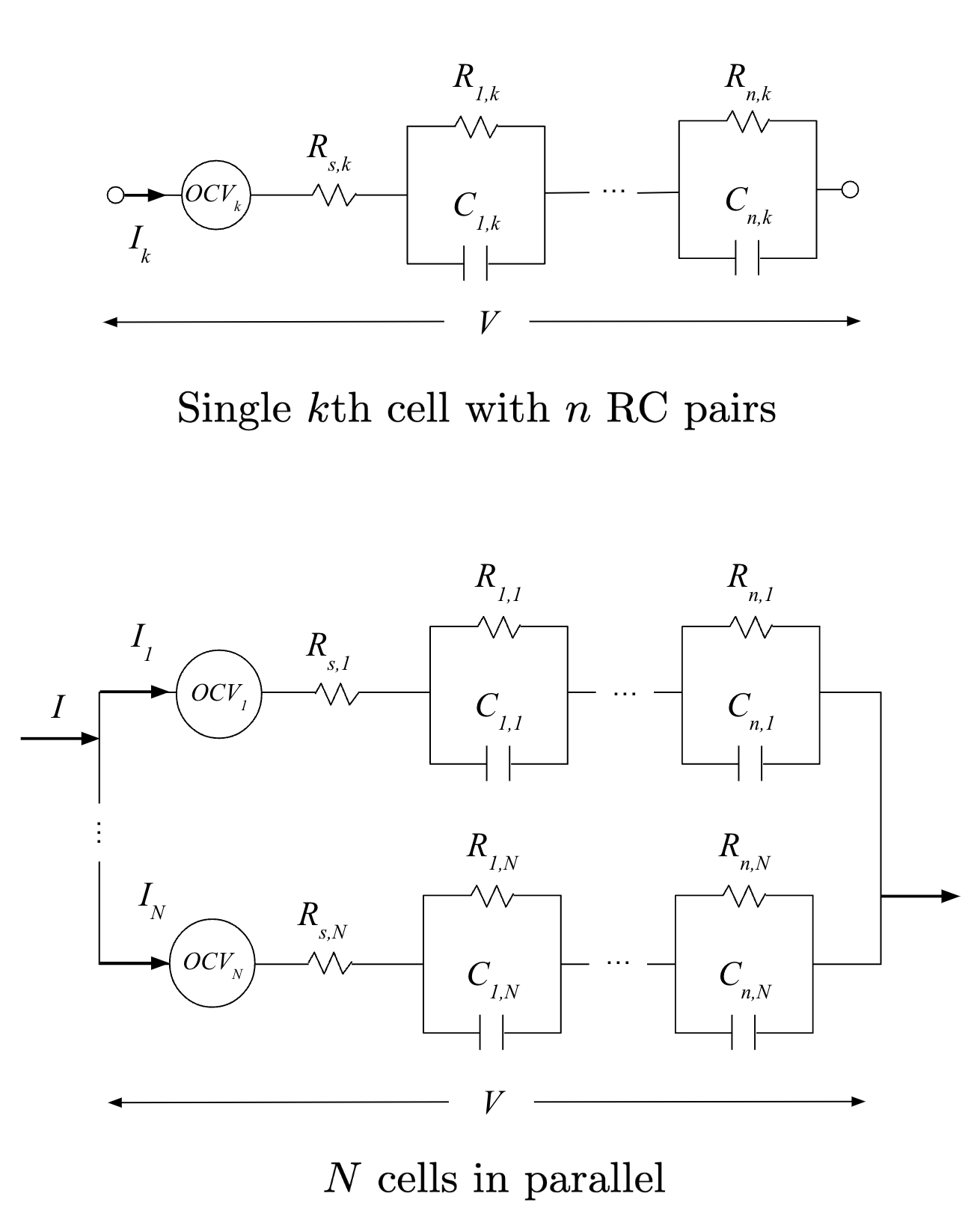}
\caption{Equivalent circuit models of a single cell and cells in parallel.}
\label{fig:circuitModels}
\end{figure}

In a forward causality battery pack model, the current, $I(t)$, is treated as an input variable, and the voltage, $V(t)$, is treated as an output variable. Different levels of model fidelity and complexity exist in the literature. Fig.~\ref{fig:circuitModels} shows one example of a nonlinear equivalent circuit model (ECM) of a single battery cell. This model contains three elements: \textit{(i)} a nonlinear relationship between OCV and SOC, \textit{(ii)} an Ohmic series resistor $R_s$, and \textit{(iii)} a series connection of $n$ resistor-capacitor (RC) pairs. These RC pairs, with resistances $R_{1,..,n}$ and capacitances $C_{1,...,n}$, represent the dynamics associated with ionic diffusion, double-layer effects, etc. Mathematically, this model leads to the following state-space representation: 

\begin{equation}
    \begin{split}
        \dot{x}_{1,k}(t) &= \frac{1}{Q_k} I_k(t) \\
        \dot{x}_{2,k}(t) &= I_k(t) - \frac{x_{2,k}(t)}{R_{1,k}C_{1,k}}\\
        &... \\
        \dot{x}_{n+1,k}(t) &=  I_k(t) - \frac{x_{n+1,k}(t)}{R_{n,k}C_{n,k}} \\
        V_k(t) &= g_k(x_{1,k}) + I_k(t)R_{s,k} + \hdots + \frac{x_{n+1,k}}{C_{n,k}}
    \end{split}  \label{eq:forwardStateEquations}
\end{equation}

Here, the integer $k \in \{1,2,...,N\}$ is the index of an individual cell in a set of $N$ parallel cells. The portion of the total input current through this cell is $I_k$. The first state variable, $x_{1,k}$, represents the cell's SOC. The remaining state variables $x_{2,...,n+1,k}$ represent the amounts of charge stored in the capacitors $C_{1,...,n,k}$, respectively. Finally, the function $g_k(x_{1,k})$ represents the nonlinear OCV-SOC relationship for the given cell. 

Parallel-connected cells share the same output voltage and split the input current. This leads to the following algebraic constraints: 
\begin{equation}
    \begin{split}
        I(t) &= I_1(t) + I_2(t) + ... + I_N(t) \\
        V_1(t) & = V_2(t) = ... = V_N(t)
    \end{split}
\end{equation}

Altogether, the above equations translate to a DAE model: the differential equations represent the dynamics of individual cells, and the algebraic constraints represent Kirchoff's laws for parallel connections.

\subsection{Inverse Causality Model}
\label{subsec3}

The concept of \textit{causality} in the control systems literature refers to the choice of assigning variables as the inputs versus outputs of a given dynamic system~\cite{karnopp2012system}. Some dynamic systems have strict causality, meaning it can only be expressed in state-space form with one choice of input and output variables. In contrast, the ECM in Fig.~\ref{fig:circuitModels} is not strictly causal because the Ohmic series resistance, $R_s$, introduces a \textit{feedthrough effect}---an instantaneous relationship between a cell's current and voltage. As long as $R_s$ is nonzero, one can invert the causality of the battery model to obtain a new state-space representation with voltage as the input variable and current as the output variable. This simply means that we rewrite forward causality model's output equation as follows: 

\begin{equation}
    I_k = \frac{V_k(t) - g_k(x_{1,k}) -  \frac{x_{2,k}}{C_{1,k}} - \hdots - \frac{x_{n+1,k}}{C_{n,k}}}{R_{s,k}}
\end{equation}

This relationship represents a new \textit{output} equation for the $k$th cell, where $V_k$ is now treated as its input variable and $I_k$ is treated as its output variable. Plugging this relationship into the $k$th cell's state equations gives: 

\begin{equation}
    \begin{split}
        \dot{x}_{1,k}(t) &= \frac{1}{Q_k} \frac{V_k(t) - g_k(x_{1,k}) -  \sum_{j=1}^{n}\frac{x_{j+1,k}}{C_{i,k}}}{R_{s,k}} \\
        \dot{x}_{j,k}(t) &= \frac{V_k(t) - g_k(x_{1,k}) -  \sum_{j=1}^{n}\frac{x_{j+1,k}}{C_{i,k}}}{R_{s,k}} \\ 
        &- \frac{x_{j,k}(t)}{R_{j-1,k}C_{j-1,k}}, \\
    \end{split}
    \label{eq:inverseStateEquations}
\end{equation}

\noindent where $j \in \{2,...,n+1\}$ is used for indexing the state variables associated with RC pair dynamics. 

Since the voltages $V_{1,2,...,N}$ of parallel-connected cells are equal, one can replace $V_k$ in the above state-space model with $V$, the terminal voltage across all parallel cells. Doing so furnishes a cell-by-cell state-space model consisting of \textit{explicit} ODEs, with no algebraic constraints. In other words, causality inversion breaks the algebraic loop in the parallel-connected pack model, thereby converting it from a set of DAEs to ODEs. The output of this model is total pack current, $I(t)$, related to individual cell currents and ultimately to the model's state equations as follows: 

\begin{equation}
    \begin{split}
        I(t) &= \sum_{k=1}^{N}I_k(t) \\
        &= \sum_{k=1}^{N}\frac{V(t) - g_k(x_{1,k}) -  \sum_{j=1}^{n}\frac{x_{j+1,k}}{C_{i,k}}}{R_{s,k}}
    \end{split} 
    \label{inverseOutputEquation}
\end{equation}

Intuitively, the above causality inversion process simply uses the nonzero Ohmic resistance of each cell to reformulate its state-space dynamics with voltage as the input variable and current as the output variable. The reformulation is exact: it involves no model reduction or simplification. The forward causality model makes it possible to simulate a battery pack's dynamics versus time, given its initial conditions and input current versus time. The inverse causality model, in contrast, makes it possible to simulate the same dynamics versus time, given the same initial conditions plus input voltage versus time. Both models are equally valid for use in online state estimation using algorithms such as Kalman filtering, because these algorithms only require the availability of both input and output measurements versus time. Please note that the choice of model causality is important for \textit{online control} applications, where a BMS needs to dictate the physical input variable, given measurements of the physical output variable. However, this choice is immaterial for \textit{online estimation}, including SOC estimation. Ultimately, the beauty of the above inverse causality modeling approach is the degree to which it both simplifies observability analysis and enables computationally efficient online state estimation, as shown in the following sections. 


\section{Observability Analysis}
\label{sec3}

Given a dynamic model, control theory states that the model is \textit{observable} if its internal state variables can be estimated from measurements of both its input and output variables versus time~\cite{friedland2012control}. Two types of observability analysis exist in the literature---linear and nonlinear. If a battery model satisfies linear observability conditions, then the state variables of a locally linearized model of the battery can be estimated from input-output measurement data. This makes linear observability a desirable property in battery state estimation. However, even in the absence of linear observability, it may still be possible to exploit specific features of nonlinear battery behavior for state estimation, such as the curvature of a cell's OCV-SOC relationship, as well as the possible dependence of a cell's Ohmic resistance on SOC. Nonlinear observability analysis provides a more comprehensive perspective on the feasibility of battery state estimation, encompassing such special scenarios~\cite{xu2023improving,allam2021linearized}. This paper focuses on analyzing the linear observability of parallel-connected battery cells, leaving nonlinear observability as an open topic for potential future exploration. We begin by linearizing the inverse causality battery model, then analyze its observability for the special case where RC dynamics are neglected, then finally generalize the analysis to incorporate an arbitrary number of RC pairs. 

\subsection{Model Linearization}
\label{subsec4}

This section linearizes the inverse causality battery model around an arbitrary equilibrium. By definition, at equilibrium, the rates of change of all state variables must equal zero. This definition of an equilibrium applies to both the forward and inverse causality models. Applying this definition to the forward causality model in Eq.~\ref{eq:forwardStateEquations} leads to the conclusion that at equilibrium: 

\begin{itemize}
    \item The current, $I_k$, must equal zero. 
    \item The state of charge, $x_{1,k}$, is arbitrary. 
    \item All other state variables, $x_{2,...,n+1,k}$, must be zero. 
\end{itemize}

Linearizing a state-space model around an equilibrium involves computing the Jacobians of both its state and output equations around it. Applying this to the inverse causality model in Eq.~\ref{eq:inverseStateEquations} and Eq.~\ref{inverseOutputEquation} leads to the following linearized state-space model with inverse causality: 
\begin{equation}
    \begin{split}
        \dot{x}_{1,k}(t) &= \frac{1}{Q_k} \frac{V(t) - \gamma_k x_{1,k} -  \sum_{j=1}^{n}\frac{x_{j+1,k}}{C_{i,k}}}{R_{s,k}} \\
        \dot{x}_{j,k}(t) &= \frac{V(t) - \gamma_k x_{1,k} -  \sum_{j=1}^{n}\frac{x_{j+1,k}}{C_{i,k}}}{R_{s,k}} \\ 
        &- \frac{x_{j,k}(t)}{R_{j-1,k}C_{j-1,k}} \\
        I(t) &= \sum_{k=1}^{N}\frac{V(t) - \gamma_k x_{1,k} -  \sum_{j=1}^{n}\frac{x_{j+1,k}}{C_{i,k}}}{R_{s,k}},
    \end{split}
    \label{eq:linearizedModel}
\end{equation}

\noindent where the state, input, and output variables in the above model represent perturbations from equilibrium, and $\gamma_k$ is the local slope of each battery cell's OCV-SOC relationship around its equilibrium SOC. Note that the equilibrium SOCs for different cells can, but do not have to be, equal. In fact, variations in OCV-SOC functions across cells is an example of real-life pack heterogeneity.

\subsection{Observability Analysis without RC Dynamics}
\label{subsec5}

Consider a simplified 1st-order version of the linearized model in Eq.~\ref{eq:linearizedModel}, where $n=0$. Such a model captures the linearized OCV-SOC relationship as well as the instantaneous effects of the cells' Ohmic resistances, but neglects higher-order RC dynamics. This simplification leads to the following linearized state-space representation: 
\begin{equation}
    \begin{split}
        \dot{x}_{1,1}(t) &= \frac{1}{Q_1} \frac{V(t) - \gamma_1 x_{1,1} }{R_{s,1}} \\
        &...\\
        \dot{x}_{1,N}(t) &= \frac{1}{Q_N} \frac{V(t) - \gamma_N x_{1,N}}{R_{s,N}} \\
        I(t) &= \sum_{k=1}^{N}\frac{V(t) - \gamma_k x_{1,k}}{R_{s,k}}
    \end{split}
    \label{eq:linearizedModel}
\end{equation}

Suppose all the SOCs of the individual battery cells are grouped into a vector, $\mathbf{z}(t)$, as follows: 
\begin{equation}
    \mathbf{z}(t) = [x_{1,1}(t),~...,~x_{1,N}(t)]^T
\end{equation}

Given the above state vector definition, one can rewrite the above state-space model in matrix form as follows: 
\begin{equation}
    \dot{\mathbf{z}} = \mathbf{A}\mathbf{z}(t) + \mathbf{B}V(t),~~I(t) = \mathbf{C}\mathbf{z}(t) + \mathbf{D}V(t), 
    \label{ssMatrices1}
\end{equation}

\noindent where the state, input, output, and feedthrough matrices are given by: 
\begin{equation}
    \begin{split}
        \mathbf{A} &= \begin{bmatrix}
        -\frac{\gamma_1}{Q_1R_{s,1}} & \hdots & 0 \\ 
        \vdots & \ddots & \vdots \\ 
        0 & \hdots & -\frac{\gamma_N}{Q_NR_{s,N}}\end{bmatrix}, \quad
        \mathbf{B} = \begin{bmatrix}\frac{1}{Q_1R_{s,1}} \\ \vdots \\ \frac{1}{Q_NR_{s,N}  }\end{bmatrix}, \\ \mathbf{C} &= \begin{bmatrix}-\frac{\gamma_1}{R_{s,1}} & ... & -\frac{\gamma_N}{R_{s,N}}\end{bmatrix}, \quad \mathbf{D} = \frac{1}{R_{s,1}} + ... + \frac{1}{R_{s,N}}
    \end{split}
    \label{eq:ssMatrices2}
\end{equation}

The above linearized model is observable if and only if the observability test matrix $\mathcal{O}_N$ below has full column rank~\cite{friedland2012control}. 
\begin{equation}
\mathcal{O}_N = 
\begin{bmatrix}
\mathbf{C} \\ \mathbf{CA} \\ \mathbf{CA}^2 \\ \vdots \\ \mathbf{CA}^{N-1} \\
\end{bmatrix}.
\end{equation}

Expanding the above observability test matrix gives: 
\begin{equation}
    \mathcal{O}_{N} = 
    \begin{bmatrix}
    -\frac{\gamma_1}{R_s,1} &  -\frac{\gamma_2}{R_s,2} & \hdots & -\frac{\gamma_N}{R_s,N} \\
    \frac{\gamma_1^2}{Q_1R_{s,1}^2} &  \frac{\gamma_2^2}{Q_2R_{s,2}^2} & \hdots & \frac{\gamma_N^2}{Q_NR_{s,N}} \\
    \vdots & \vdots & \ddots & \vdots \\
    \frac{(-1)^N\gamma_1^N}{Q_1^NR_{s,1}^N} & \frac{(-1)^N\gamma_2^N}{Q_2^NR_{s,2}^N} & \hdots & \frac{(-1)^N\gamma_N^N}{Q_N^NR_{s,N}^N}
    \end{bmatrix} 
\end{equation}

\noindent The columns of this observability test matrix can be written as 

\begin{equation}
    -\frac{\gamma_1}{R_{s,1}}
    \begin{bmatrix}
     1 \\ \lambda_1 \\ \lambda_1^2 \\ \vdots \\ \lambda_1^N
    \end{bmatrix}, 
    -\frac{\gamma_2}{R_{2,s}}
    \begin{bmatrix}
     1 \\ \lambda_2 \\ \lambda_2^2 \\ \vdots \\ \lambda_2^N
    \end{bmatrix}, 
     - \hdots, -
     \frac{\gamma_N}{R_{s,N}}
    \begin{bmatrix}
     1 \\ \lambda_N \\ \lambda_N^2 \\ \vdots \\ \lambda_N^N
    \end{bmatrix},
    \label{eq:columns}
\end{equation}

\noindent where the symbol $\lambda_k = -\frac{\gamma_k}{Q_kR_{s,k}}$, $k \in \{1, 2, \hdots, N\}$ denotes the eigenvalue associated with the dynamics of the $k$th cell, which has an important intuitive meaning: the absolute values of its reciprocal is the time constant associated with the exponential relaxation of each cell's charge/discharge current during potentiostatic (i.e., constant voltage) operation. 

Given the above observability test matrix, the main result of this paper can be stated as follows:

\newtheorem{theorem}{Theorem}
\begin{theorem}
    The observability of the state-space model in Eq.~\ref{eq:ssMatrices2} has the necessary and sufficient conditions that: \textit{(i)} the slopes of all cells' characteristic OCV-SOC curves, $\gamma_{1,...,N}$, are nonzero; \textit{(ii)} the series Ohmic resistances of all cells, $R_{1...s,N}$ are finite; and \textit{(iii)} the potentiostatic current relaxation eigenvalues of all cells $\lambda_{1,...,N}$ are distinct.
\end{theorem}

\begin{proof}

We begin by noting that if any OCV-SOC slopes, $\gamma_k$, equals zero or if any of the series Ohmic resistances, $R_{s,k}$, is infinite, then its corresponding column in the observability test matrix will equal zero. The observability test matrix will be rank deficient, and observability will be lost. Therefore, it is a necessary condition for observability that all OCV-SOC slopes be nonzero, and that all series resistances be finite.

Assuming this condition is met, we note that the columns of the observability test matrix, regardless of any nonzero scalars multiplying them, must be linearly independent. We therefore assemble the columns in \ref{eq:columns}, without their (nonzero) pre-multipliers, into a new matrix that represents an observable system if and only if its determinant is nonzero:

\begin{equation}
\mathcal{O}_N^* = 
\begin{bmatrix}
    1 & 1 & \hdots & 1 \\
    \lambda_1 & \lambda_2 & \hdots & \lambda_N \\
    \lambda_1^2 & \lambda_2^2 & \hdots & \lambda_N^2 \\
    \vdots & \vdots & \ddots & \vdots \\
    \lambda_1^N & \lambda_2^N & \hdots & \lambda_N^N \\
\end{bmatrix}
\end{equation}

This happens to be a Vandermonde matrix. Its columns consist geometric sequencies of $N$ variables, in this case $\lambda_k$, from the $0$th to the $N$th order. The Vandermonde determinant is given by

\begin{equation}
\det(\mathcal{O}_N^*) = \prod (\lambda_i - \lambda_j), 0 \leq i < j \leq N,
\end{equation}

\noindent which is nonzero \textit{if and only if all $N$ variables, $\lambda_k$, are distinct.} This gives us another necessary condition, that the eigenvalues of all cells must be distinct. 

Therefore, in the case of $N$ cells with $n=0$ RC pairs, the SOCs of parallel cells are unobservable if any of the following statements are true: 

\begin{itemize}
    \item $\gamma_k = 0 \vee R_{s,k} = \infty$.
    \item Any two or more eigenvalues of $\mathbf{A}$ are equal. 
\end{itemize}

This proves necessity. For sufficiency, note that if the eigenvalues $\lambda_{1,...,N}$ are indeed distinct, then the  Vandermonde matrix is full rank. Moreover, any matrix whose columns are nonzero multiples of those in the Vandermonde matrix will have linearly independent columns, and will therefore be full rank. Assuming that the OCV-SOC slopes of the individual battery cells are nonzero and the series Ohmic resistances of these individual cells are finite, the observability test matrix will be full rank. This proves sufficiency. 
\end{proof}

\subsection{Observability Analysis with RC Dynamics}
\label{subsec6}

As seen above, inverse 1st-order dynamics have a diagonal state matrix $\mathbf{A}$. Hence, diagonalizing the dynamics of a single cell model with $n \neq 0$ RC pairs makes it a mathematical equivalent of $n+1$ parallel-connected 1st-order cell models. Each state---both SOCs and capacitor charges---can then be reinterpreted as the SOC of a 1st-order cell, such that we lose the mathematical (but not numerical) distinction between the two models. Extending to any number of  $N$ physical cells requires only the extension of this model in that manner. The observability analysis and results would then be identical to the $n=0$ case, meaning it is generalizable to any number of $N$ cells each containing any number of $n$ RC pairs. 

These results demonstrate that causality inversion can significantly simplify observability analysis. Intuitively, parts \textit{(i)} and \textit{(ii)} of Theorem 1 state that each cell must be capable of being charged and discharged and of experiencing a voltage change as a result. Moreover, part \textit{(iii)} states that the dynamics of individual battery cells must be distinguishable from one another. The latter implies that grouping identical cells into one fictitious larger cell will restore observability, as we demonstrate with our clustering criterion in the next sections. 


\section{Kalman Filter}
\label{sec4}

This section applies the above mathematical insights to the design of a Kalman filter for SOC estimation in parallel-connected cells. The section begins by presenting two ECMs of different battery cells, based on laboratory chracterization experiments. A clustering criterion is then developed for grouping battery cells connected in parallel to ensure SOC observability. Finally, the section presents a Kalman filter based on this clustering criterion plus inverse dynamic modeling.

\subsection{Model Identification}
\label{subsec7}

\begin{figure}[h]
\centering
\includegraphics[scale=0.55]{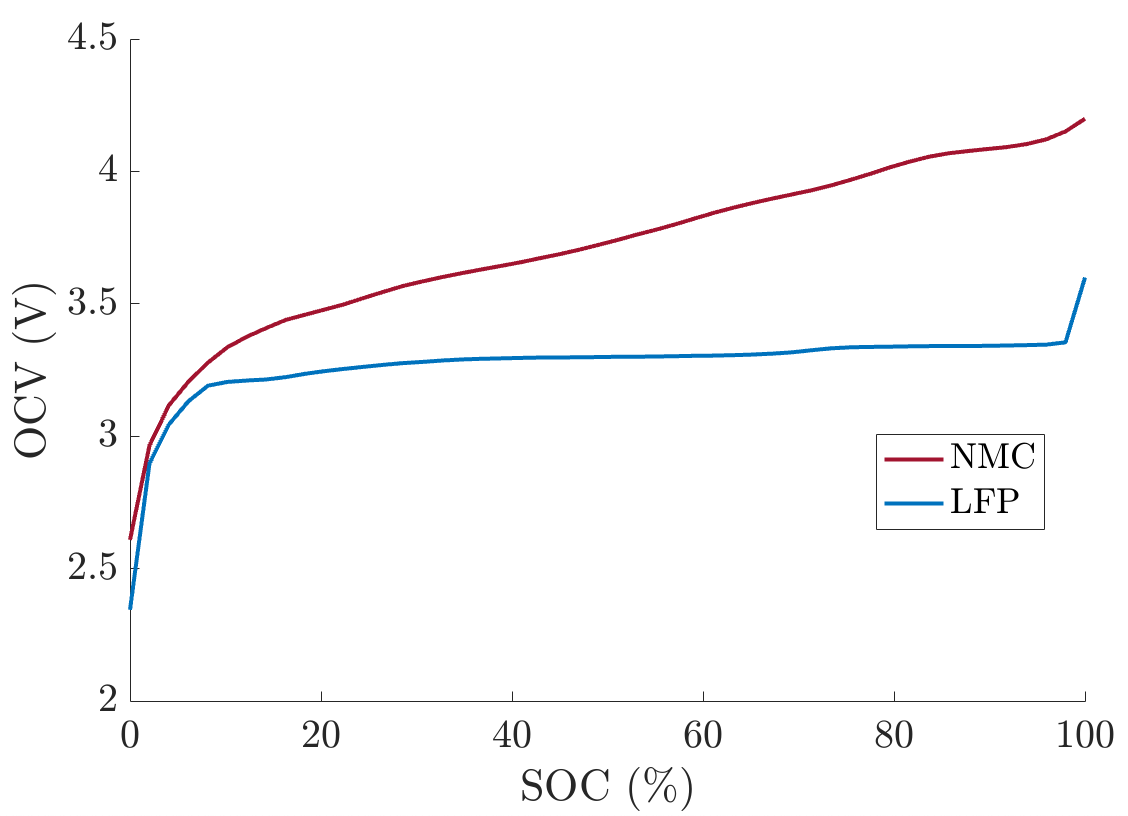}
\caption{CCCV cycling results for OCV as a function of SOC.}
\label{fig:ocvSOC}
\end{figure}

Two different commercial battery cell chemistries are used to demonstrate the findings of this paper. These are a Molicel 18650 2800 mAh nickel-manganese cobalt (NMC) cell and a LithiumWerks 18650 1100 mAh lithium iron phosphate (LFP) cell. We characterized both cells experimentally using an Arbin BT-2000 cycler. Specifically, slow constant current, constant voltage (CCCV) cycling was used to characterize the two cells' charge capacities $Q$ and OCV-SOC curves, while faster pulse testing was used to determine their internal resistance $R_s$ and RC pair parameters $R_1$, $C_1$, $R_2$, and $C_2$. The two cells' ECM parameters and OCV-SOC relationships are shown in Table \ref{table:parameters} and Fig. \ref{fig:ocvSOC}, respectively.

\begin{table}[h]
\centering
\begin{tabular}{c | c | c | c | c | c | c } 
   & $Q$ (C) & $R_s$ ($\Omega$) & $R_1$ ($\Omega$) & $C_1$ (F) & $R_2$ ($\Omega$) & $C_2$ (F)\\ [1ex]  
 \hline 
 NMC & 9925 & 0.102 & 9.4e-3 & 6330 & 3.63e-2 & 6797\\ [1ex]  
 LFP & 4579 & 0.261 &  0.187 & 8232 & 1.75e-2 & 1749\\
 \end{tabular}
\caption{ECM parameters up to the third order.}
\label{table:parameters}
\end{table}

\pagebreak
\subsection{Clustering Criterion}
\label{subsec8}

In Section~\ref{sec3}, we proved that the system loses observability if any two of its states exhibit identical dynamics. More specifically, observability is lost if any two cells have identical eigenvalues associated with current relaxation during potentiostatic discharge. This leads to a very simple clustering criterion. Consider a number of battery cells in parallel. Let $\lambda_k = -\frac{\gamma_k}{Q_kR_{s,k}}$, $k \in \{1, 2, \hdots, N\}$ be the eigenvalues associated with current relaxation during potentiostatic charge/discharge for each of these cells. Then grouping cells with similar eigenvalues into a single equivalent cell will ensure collective SOC observability for these parallel-connected cells. The parameters of each fictitious equivalent cell (representing a cluster of cells) are then equal to the sums of the parameters of the underlying individual cells:

\begin{align}
Q_{cluster} &= \sum Q_{individual} \\
\frac{1}{R_{s,cluster}} &= \sum \frac{1}{R_{s,individual}}
\end{align}

\begin{figure}[h]
\centering
\includegraphics[scale=0.6]{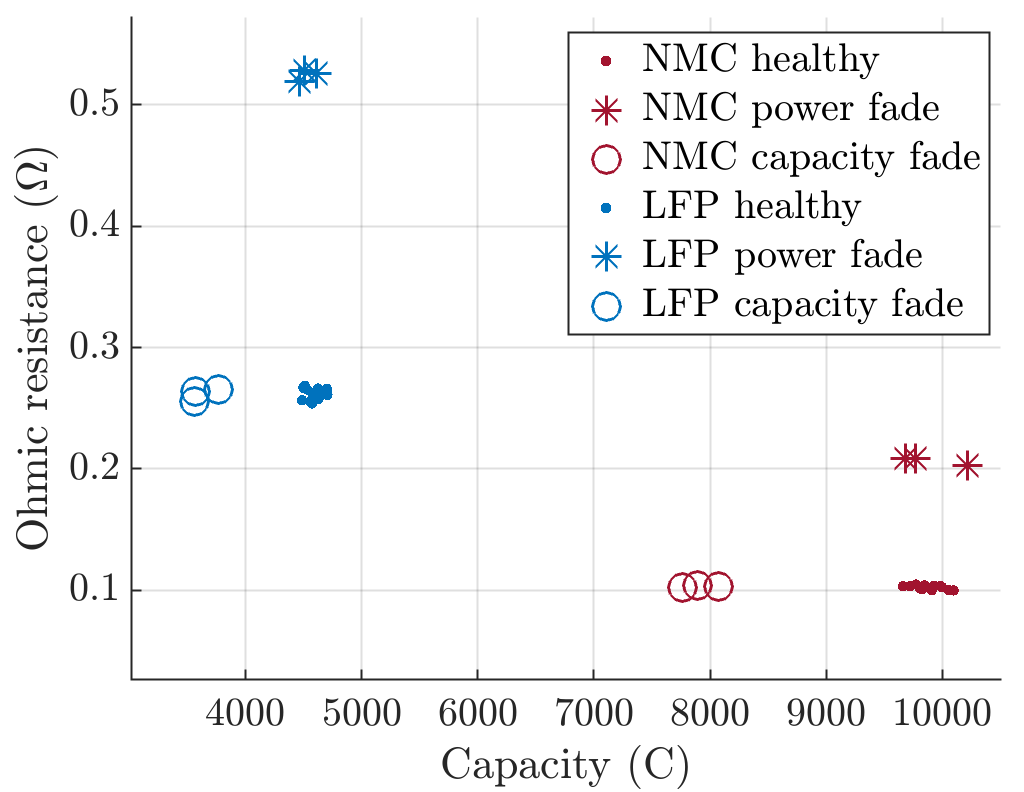}
\caption{NMC and LFP cell clusters, capacity vs. Ohmic resistance.}
\label{fig:clusters}
\end{figure}

To illustrate this clustering approach, we generate 20 fictitious cells of each battery chemistry. All 20 cells share the same OCV-SOC curve, and therefore the same slope, $\gamma$, of this curve around any nominal SOC used for model linearization. However, the cells have different 1st-order model parameters, as show in Fig.~\ref{fig:clusters}. Specifically, for each battery chemistry, we generate three groups of cells. The first group contains 14 healthy cells whose parameters are uniformly distributed within $\pm 3\%$ of the nominal parameters estimated experimentally. The second group contains 3 cells with parameters uniformly distributed in the same manner around a nominal set of values, but the nominal internal resistance is double that of the healthy cells. The third group contains 3 cells with parameters uniformly distributed in the same manner around a nominal set of values, but with nominal charge capacity that is only $80\%$ of the nominal healthy cell capacity. In a practical battery pack, the second and third groups represent cells with significant power fade and significant capacity fade, respectively. This distribution of battery cells translates to 3 clusters for each battery chemistry, based on our clustering criterion: (i) a healthy cell cluster packed around a healthy current relaxation eigenvalue; (ii) a cluster of cells suffering from power fade and packed around 0.5 times this eigenvalue; and (iii) a cluster of cells suffering from capacity fade and packed around 1.25 times this eigenvalue. 

\subsection{Design of Kalman Filter}
\label{subsec9}

The Kalman filter, used widely in battery state estimation, is an optimal state estimator that minimizes estimation error given a system's characteristic process and measurement noise. In our simulation, we implement a steady-state corrective Kalman gain matrix $\mathbf{L}$. The nonlinear OCV-SOC relationships of both the NMC and LFP cells mean that the parameter $\gamma_k$ in each $k$th cell varies nontrivially across a charge cycle. We circumvent the resulting need for a time-varying filter model via a single linearization as described in Section~\ref{sec3}: the $\gamma_k$ terms in $\mathbf{L}$ are calculated as the slope between 40\% and 60\% SOC in the above OCV-SOC plots. A time-varying model would require the computation of $\mathbf{L}$, which involves solving the algebraic Riccati equation, at every or several timesteps. Our assumption of a time-invariant system is intended to minimize the computational resources required of a BMS.

Based on the matrices $\mathbf{A,B,C,D}$ from the inverse causality state-space model in \ref{ssMatrices1} and \ref{eq:ssMatrices2}, the continuous-time steady-state Kalman filter recursively predicts and updates the state values based on the following equations: 

\begin{align}
\mathbf{P}(t) &= \mathbb{E}[(\mathbf{x}(t)-\mathbf{\hat{x}}(t))(\mathbf{x}(t)-\mathbf{\hat{x}}(t))^T], \mathbf{Q} = \mathbb{E}[w^2], \mathbf{R} = \mathbb{E}[v^2] \\
\frac{d}{dt}\mathbf{P}(t\rightarrow\infty) &= \mathbf{AP} + \mathbf{PA}^T - \mathbf{PC}^T\mathbf{R}^{-1}\mathbf{CP} + \mathbf{BQB}^T = 0\\
\mathbf{L} &= \mathbf{L}(t\rightarrow\infty) = \mathbf{PC}^{T}\mathbf{R}^{-1} \\
\hat{y}(t) &= \mathbf{C}\mathbf{\hat{x}}(t) + \mathbf{D}u(t)\\
\frac{d\mathbf{\hat{x}}(t)}{dt} &= \mathbf{A}\mathbf{\hat{x}}(t) + \mathbf{B}u(t) + \mathbf{L}(y(t) - \hat{y}(t))
\end{align}

\noindent where $u$ and $y$ represent, in the inverse causality model, the scalar input (total pack voltage) and scalar output (total pack current). Hats indicate estimates. Crucially, as explained in Section~\ref{sec2}, causality inversion enables us to implement this Kalman filter without DAEs, which reduces the its complexity. The symbol $w$ represents the additive Gaussian process or input measurement noise, and $v$ represents the additive Gaussian output measurement noise. In our case, $w$ and $v$ are assumed not to be correlated. In the following simulations we set the process noise magnitude to $500$ $\mu$V and measurement noise magnitude to $20$ mA.

\section{Monte Carlo Simulation Studies}
\label{sec5}

This section demonstrates the previous mathematical insights and filter design approach using four Monte Carlo simulation studies. All four simulation studies utilize 1st-order models (with no RC pairs) as foundations for SOC estimation via Kalman filtering. However, in two of the simulation studies, the underlying battery pack dynamics are simulated using 3rd-order models. This is more representative of battery pack SOC estimation in real life, where estimation algorithms (e.g., Kalman filters) are typically designed using reduced-order dynamic models of physical batteries with higher-order dynamics. The end product is a set of four simulation studies for: (i) an NMC pack with 20 1st-order ECMs; (ii) an NMC pack with 20 3rd-order ECMs; (iii) an LFP pack with 20 1st-order ECMs; and (iv) an LFP pack with 20 3rd-order ECMs.

\subsection{Simulation of Pack}
\label{subsec10}

\begin{figure}[h]
\centering
\includegraphics[scale=0.4]{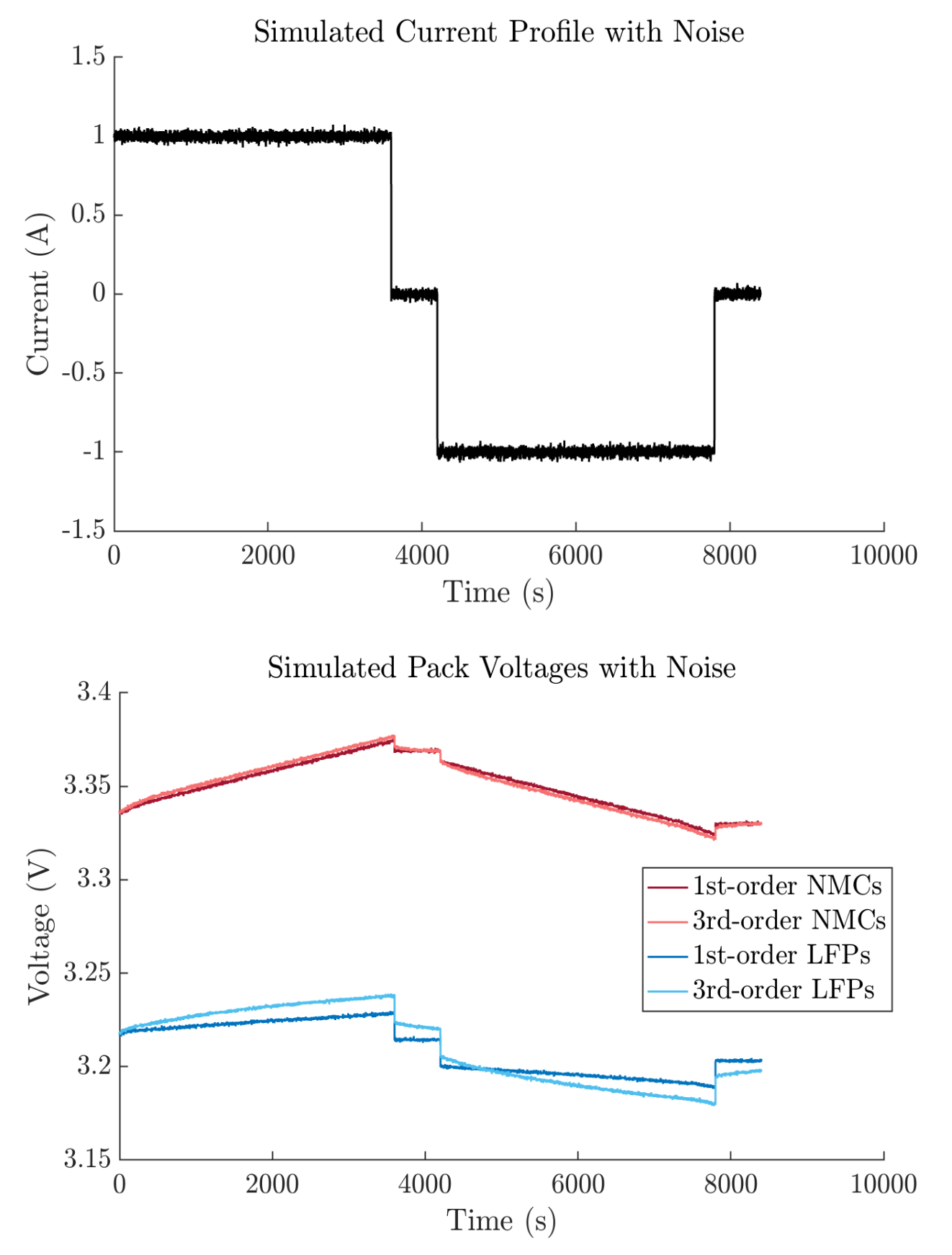}
\caption{Simulated current and voltage for 4 packs with 1st or 3rd-order NMC or LFP cells.}
\label{fig:simulatedPacks}
\end{figure}

Each of the four studies contains 100 simulations of charge and discharge for the same NMC or LFP pack with uniformly distributed parameters. The parameters associated with the RC pairs in the 3rd-order ECMs were kept constant. All 400 simulations have a simple current profile of 1 A charge and discharge for 1 hour with 10 minute rests in between, and a true initial SOC of 10\% across all 20 cells. The simulated noisy pack current profile and examples of noisy pack voltage profiles for NMC and LFP are shown in Fig.~\ref{fig:simulatedPacks}. 

\subsection{Simulation of Kalman Filter}
\label{subsec12}

All 400 simulations apply the steady-state Kalman filter that relies on the 1st-order inverse causality model. Each simulation has a different initial SOC estimate uniformly distributed across the entire SOC range of 0 to 1. Pseudo-random Gaussian noise was added to the simulated current and voltage measurements using Matlab's \textit{randn} function. 

\begin{figure}[h]
\centering
\includegraphics[scale=0.31]{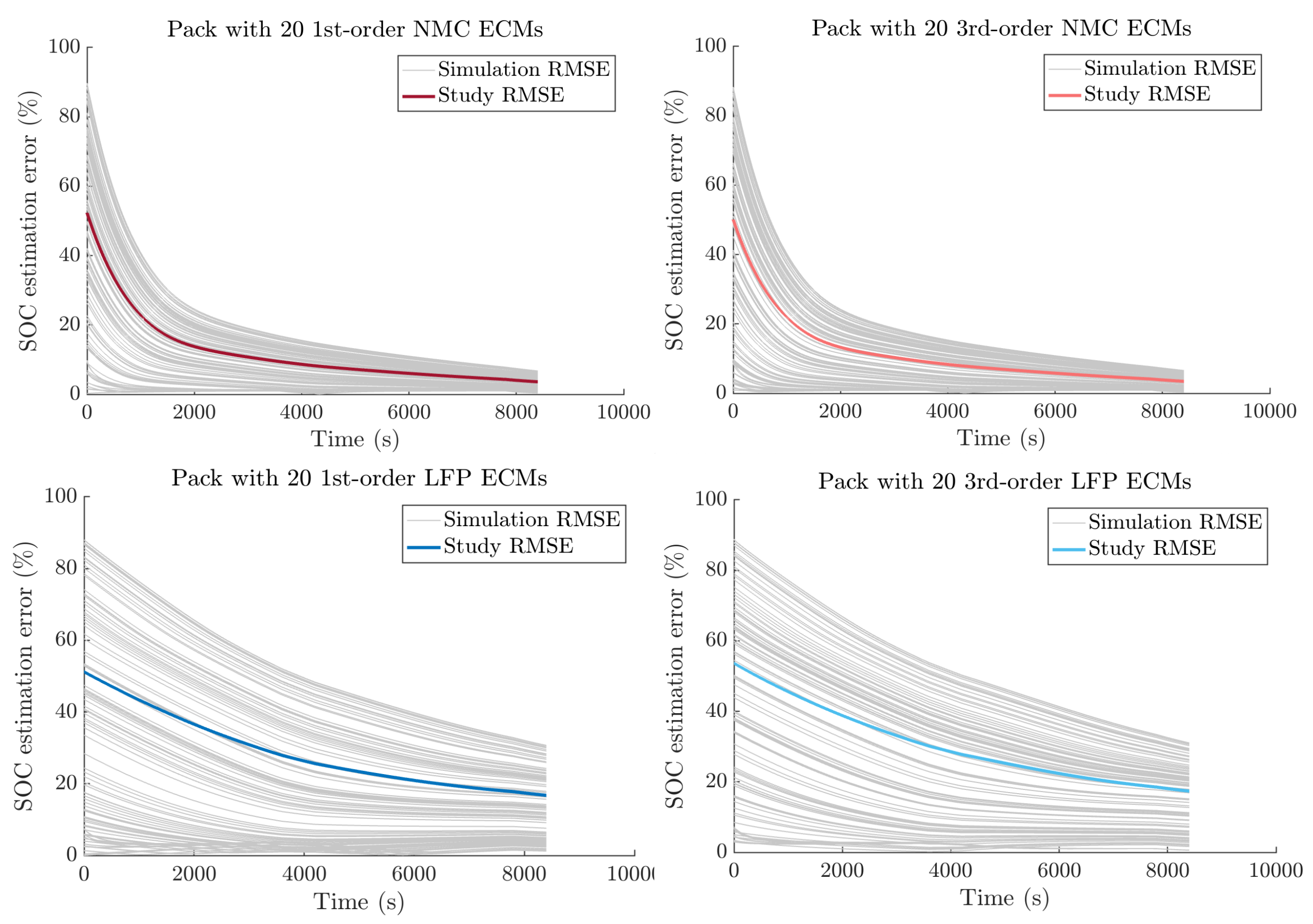}
\caption{Error vs. time of the four Monte Carlo studies.}
\label{fig:eachStudy}
\end{figure}

\begin{figure}[h]
\centering
\includegraphics[scale=0.55]{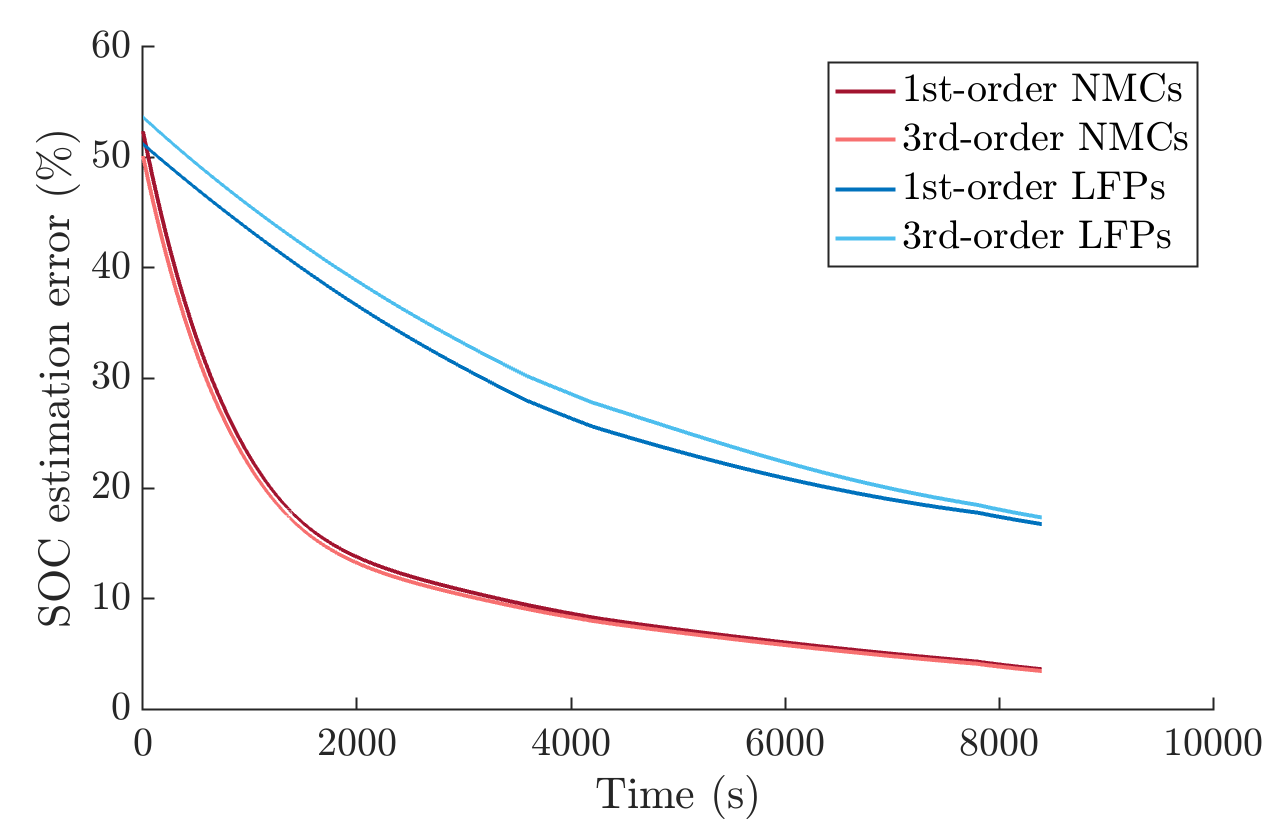}
\caption{Overall RMSEs of the four Monte Carlo experiments.}
\label{fig:overallStudy}
\end{figure}

As shown in Fig.~\ref{fig:overallStudy} and~\ref{fig:eachStudy}, the 1st-order steady-state Kalman filter's overall study errors and individual simulation errors show converging behavior for both the 1st-order and 3rd-order ECM pack plants regardless of the initial estimate error, though more slowly for LFPs. This is consistent with the well-established fact that fast and accurate SOC estimation can be quite challenging for the LFP chemistry, given the flatness of its OCV-SOC curve over a broad range of SOCs. 

Perhaps the most exciting aspect of the above estimation results is their the fact that they show good convergence properties, consistent with the literature, \textit{without the need for solving a DAE model}. This is a direct consequence of inverse causality modeling, and is quite an appealing research outcome because of its computational feasibility and algorithm scalability. Moreover, the fact that the algorithm's SOC estimates do converge is consistent with the fact that the underlying battery pack model used for SOC estimation is observable. This, in turn, is a direct outcome of the fact that this pack-level model is obtained by clustering different battery cell dynamics in a manner that theoretically ensures linear observability. 

\section{Conclusion}
The central conclusion of this paper is that inverse dynamic modeling significantly simplifies observability analysis and estimation for the SOCs of cells in parallel-connected battery packs. This is supported by mathematical analysis and simulated Monte Carlo studies of a Kalman filter that relies on causality inversion and a simple clustering criterion. The significance of our conclusion is threefold. First, it presents a very simple mathematical reformulation that eliminates the computational complexity associated with DAEs for application in real-time BMSs. Second, our simple addition of a clustering criterion to the estimation algorithm enables SOC estimation for battery packs with cell-to-cell variability while preserving the system's observability. Third, our approach can be extended to larger battery packs, such as those with parallel modules in series, or to more complex filtering algorithms as desired.

\section*{Acknowledgments}
We would like to acknowledge the support of NSF Grant DMR2149982, REU/RET Site: Summer Research Experiences in Renewable and Sustainable Energy Technology (ReSET), the Maryland Energy Innovation Institute (MEI$^{2}$) and the University of Maryland Departments of Mechanical Engineering and Materials Science and Engineering.

\bibliographystyle{plain}
\bibliography{References}

\end{document}